\def\br{{\bf r}}
\def\bk{{\bf k}}
\def\bG{{\bf G}}

\documentstyle[prb,aps,eqsecnum,twocolumn]{revtex}

\begin{document}
\twocolumn[\hsize\textwidth\columnwidth\hsize\csname @twocolumnfalse\endcsname

\title{The charge density of semiconductors in the $GW$ approximation} 
\author{Martin M. Rieger$^*$ and R. W. Godby$^\dagger$}
\address{$^*$Cavendish Laboratory, University of Cambridge, Madingley Road,
Cambridge, CB3 0HE, UK}
\address{$^\dagger$Department of Physics, University of York, Heslington,
York YO1 5DD, UK}

\date{\today}
\maketitle

\begin{abstract}
We present a method to calculate the electronic charge density of periodic
solids in the $GW$ approximation, using the space-time method. We investigate
for the examples of silicon and germanium to what extent the $GW$ approximation
is charge-conserving and how the charge density compares with experimental
values. We find that the $GW$ charge density is close to experiment and charge
is practically conserved. We also discuss how using a Hartree potential
consistent with the level of approximation affects the quasi-particle energies
and find that the common simplification of using the LDA Hartree potential is a
very well justified.
\end{abstract}

] 

\narrowtext

\section{Introduction} 

Amongst the established methods to calculate the electronic properties of
solids, Hedin's $GW$ approximation\cite{HED65} is notable for its unrivalled
success in predicting the energy gaps of semiconductors and insulators (e.g.
Refs. \onlinecite{HYL86,GSS88,ZHL91,RKP93}). Unlike alternative methods which
are based on density functional theory and describe ground state properties by
mapping the many-electron problem onto an effective one-electron problem, it
seeks to find a solution for the exact one-particle Green's function of the
many-electron system to which it is applied. With this knowledge, a wide range
of properties of the system under consideration can be accurately described,
among them the optical excitation spectrum, the density of states, the charge
density and the total energy.

As an exact solution of the many-electron problem remains today as elusive as
ever, any attempt to determine the Green's function has to rely on skilled
approximations. The problem of finding the exact Green's function of a system
is equivalent to finding its self-energy. In the $GW$ approximation, this
quantity is approximated by the product of $G$, the Green's function, and $W$,
the screened Coulomb interaction; hence the name.

$GW$ calculations for semiconductors so far have concentrated on the optical
excitation spectrum, the one area where $GW$ has had its most striking
successes, but other aspects of electronic structure have not been investigated
with this method for real materials. This is largely due to the prohibitively
large numerical effort. Recently, however, a new technique has been developed,
the $GW$ space-time method,\cite{RGN95} that makes $GW$ calculations much
faster and allows a precise computation of the energy dependence of the
self-energy without recourse to plasmon pole models. This opens up the way to
new applications for the $GW$ formalism.

In this paper we will concentrate on the electronic charge density\cite{CDENS}
of semiconductors, present a new technique for extracting it from the Green's
function at $GW$ level and show for the first time results of $GW$ charge
density calculations for real materials at the example of Si and Ge, for which
very careful analyses of the experimental data are available for
comparison.\cite{LZD93,LZD95} We also investigate the problem of charge
conservation and the influence of using a Hartree potential consistent with the
level of approximation on the quasi-particle spectrum.

Atomic units are used throughout unless otherwise noted.

\section{Theory}

The central equation of the $GW$ approximation is the expression for the
self-energy:
\begin{equation}
\label{GW}
\Sigma(\br,\br';\tau) = iG(\br,\br';\tau)W(\br,\br';\tau).
\end{equation}

Full self-consistency in the $GW$ approximation would mean that the Green's
function that enters into Eq.\ (\ref{GW}) were itself a solution of the
equation
\begin{eqnarray}
\left( -\frac12\nabla^2 + V_{ext}({\bf r}) + V_H^{SC}({\bf r}) 
 + \Sigma^{SC}(\br,\br';\omega)\right) \lefteqn { G(\br,\br',\omega) }
 \nonumber \\
= -i\delta(\br,\br')
\end{eqnarray}
where the superscript $SC$ for $\Sigma$ and the Hartree potential $V_H$
indicates that these quantities have been determined self-consistently. In the
case of the Hartree potential this means the charge density at $GW$ level has
to be known. Such a self-consistent $GW$ calculation has never been done for a
real system and there are indications\cite{GBH95} that self-consistency will
destroy the good agreement between the computed and experimental quasi-particle
spectrum.

In this paper we employ the usual one-iteration non-self-consistent $GW$
approximation. However, we will determine the charge density and thus the
Hartree potential self-consistently at a level where the self-energy is kept
fixed at its first-iteration value. The charge density itself is an indicator
as to whether such an approach is justified, as it has been shown by Baym and
Kadanoff\cite{BAK61,BAY62} that generally only self-consistent approximations
to the self-energy conserve particle number strictly. We will therefore monitor
carefully whether our approach violates particle number.

The traditional way to set up and solve the $GW$ equations has been to express
and compute all quantities in the reciprocal space and energy domain. This
involves sums scaling with the fourth power of the number of plane waves $N$
used to represent the wavefunctions. Rojas, Godby and Needs\cite{RGN95} have
shown recently that using the most appropriate representation at the various
stages of the computation, either real space and time or reciprocal space and
energy, and changing between representations by means of Fast Fourier
Transforms, can bring this scaling down to $N^2$, thus leading to very
significant time savings.

Another point noted by Rojas et al.\ is that a detailed calculation of all
relevant quantities along the imaginary time or imaginary energy axis is
possible without recourse to a plasmon pole or similar model. This is because
on the imaginary time axis the Green's function, the polarisability and the
self-energy are all rapidly decaying smooth functions which can easily be
represented on a regular grid. Transformation to imaginary energy via an FFT is
straightforward. If the properties of $\Sigma$ on the real energy axis are
needed, analytic continuation with the help of model functions is possible.
However, for the calculation of the charge density alone, the knowledge of the
self-energy on the imaginary energy axis only is sufficient.

Having determined the self-energy $\Sigma$ on the imaginary energy axis as
described in Ref.\ \onlinecite{RGN95}, with
\begin{equation}
\Sigma(\br,\br';i\tau) = iG_0(\br,\br';i\tau)W(\br,\br';i\tau),
\end{equation}
where $G_0$ is the Green's function at LDA level, the Green's function $G$ at
$GW$ level obeys the Dyson equation
\begin{eqnarray}
\label{DYSON}
G(i\omega) & = & G_0(i\omega) + G_0(i\omega) \times \nonumber \\
&& \times\left(\Sigma(i\omega)+\Delta V_H-V_{xc}-\epsilon_0\right)G(i\omega)
\end{eqnarray}
where $G$, $G_0$, $\Sigma$, $\Delta V_H$ and $V_{xc}$ in this notation are to
be understood as matrices in a plane-wave basis and matrix multiplication of
the factors on the right hand side is implied. $\Delta V_H$ is the change in
the Hartree potential due to the density change and has to be determined
self-consistently, $V_{xc}$ is the LDA exchange correlation potential and
$\epsilon_0$ represents the shift in the Fermi level with respect to the vacuum
in the $GW$ calculation from its value in the LDA. The use of this shift
$\epsilon_0$ was first proposed by Hedin\cite{HED65} but is often neglected in
$GW$ calculations. We use it here to introduce an element of self-consistency
into the equations by ensuring that the Fermi energy is the same before and
after applying the $GW$ correction. Eq.\ (\ref{DYSON}) can directly be solved
by matrix inversion.

The relationship between charge density $\rho$ and Green's function is given by
the equation
\begin{equation}
\rho(\br) = -\frac{2}{\pi}\int_{-\infty}^0d\omega \mbox{Im} G(\br,\br;\omega)
\end{equation}
where it is assumed that the zero of energy is chosen at the Fermi level.

Since we know the charge density at the LDA level of approximation we only need
to evaluate the charge density difference $\Delta \rho$ between the LDA and the
$GW$ result. The Green's function has poles in the second and fourth quadrant of
the imaginary plane, just infinitesimally off the real axis. Using Cauchy's
Theorem and the fact that for complex energy $z$ $\Delta G(z)$ vanishes
quadratically as $\left|z\right| \rightarrow \infty$, where $\Delta G = G -
G_0$, we know that the charge density difference can just as well be calculated
by an integration along the imaginary energy axis:
\begin{equation}
\label{DRHO}
\Delta\rho(\br) = -\frac{2}{\pi}\int_{-\infty}^0 d\omega \mbox{Re}\Delta
                  G(\br,\br;i\omega).
\end{equation}

This means that the knowledge of the Green's function along the imaginary axis
is sufficient to calculate the charge density at $GW$ level without the need
for explicit analytic continuation to the real energy axis, which would
inevitably have to involve some sort of model.

In theory, the integration of $\Delta G(i\omega)$ according to Eq.\
(\ref{DRHO}) is straightforward. In practice, in order to achieve satisfactory
convergence in a numerical integration, one would have to include such a large
number of energy points in the integration, as to make it prohibitively
expensive, because for each energy point the self-energy would first have to be
computed and the Dyson equation solved. On the other hand, we know that the
self-energy as a function of imaginary energy decays quadratically for large
energies. This allows us for large $\omega$ to make an expansion of
$G(i\omega)$ in powers of $1/\omega$ and neglect terms of order higher than
$1/\omega^2$. The perturbatively treated high-energy tail can be integrated
analytically from some energy $\omega_0$ to infinity, as we will show in the
appendix. Up to $\omega_0$ we integrate Eq.\ (\ref{DRHO}) numerically.
$\omega_0$ is thus treated as a convergence parameter.

For each new $\Delta\rho$ we compute a new $\Delta V_H$ and solve Eqs.\
(\ref{DYSON}) and (\ref{DRHO}) repeatedly until $\Delta V_H$ is stable.

\section{Results}

Before we can make comparisons to experimental results we have to account for
the fact that we work in the pseudopotential approximation. This means that
the charge density computed as described in the previous section contains no
contribution from the core electrons and the contribution of the valence
electrons is modified in the core region as well.

We follow Nielsen and Martin\cite{NIM85} in writing the total charge density as
the superposition of the atomic densities plus a deformation density
$\rho_{\mathrm{def}}$, which is defined in reciprocal space as
\begin{equation}
\rho_{\mathrm{def}}(\bG) = \rho_{\mathrm{solid,ps}}(\bG)
                         - S(\bG)\rho_{\mathrm{atom,ps}}(\bG),
\end{equation}
where $\bG$ is a reciprocal lattice vector, $S$ the structure factor of the
crystal and $\rho_{\mathrm{solid,ps}}$ the charge density generated by the
pseudowavefunctions of the valence electrons in the solid and
$\rho_{\mathrm{atom,ps}}$ the same in the free atom. The Fourier coefficients
of the total charge density in the solid are then computed according to
\begin{equation}
\label{RHOSOLID}
\rho_{\mathrm{solid}}(\bG) = S(\bG)\rho_{\mathrm{atom}}(\bG)
			 + \rho_{\mathrm{def}}(\bG),
\end{equation}
where $\rho_{\mathrm{atom}}$ is the full charge density of valence and core
electrons of a free atom. This approach assumes rigid cores, in line with the
assumptions underlying the pseudopotential method.

Where we find that the charge in the $GW$ approximation is not strictly
conserved, we normalise the density by multiplying all structure factors
computed according to Eq.\ (\ref{RHOSOLID}) by a constant that assures the
overall correct number of electrons per unit cell.

The results we present in this section come from LDA calculations that were
performed with pseudopotentials generated by the Hamann method,\cite{HAM89} a
plane-wave cutoff of 17Ry for Si and 20Ry for Ge and 10 special $\bk$ points.
For Ge relativistic effects were included in the pseudopotential but no
spin-orbit coupling was taken into account in the solid. A Ceperley-Alder
exchange correlation potential\cite{CEA80} was used in the parametrisation by
Perdew and Zunger.\cite{PEZ81} In the $GW$ space-time calculation we used 65
bands truncated to 169 plane waves from the LDA calculation. The time grid
comprised 60 points spaced at 0.314 a.u.\ which were zero-padded to 120 points
before transformation to imaginary energy, giving an $\omega_0$ of 10 Hartrees.
Other parameters of the $GW$ calculation were as described in Ref.\
\onlinecite{RGN95}.

Table \ref{FFSI} shows our computed structure factors of the pseudo valence
charge of silicon for several reciprocal lattice vectors $\bG$. The second
column shows the LDA values and the third the $GW$ correction $\Delta\rho$.
$\Delta\rho$ is actually the difference between the normalised $GW$ density
and the LDA density. Without normalisation the $GW$ valence density
integrates to a total of 8.0233 electrons per unit cell (instead of 8), a
charge violation of not quite 0.3\%. This observation of a very small but
finite charge violation is in line with rigorous analytical results by
Schindlmayr for a model system.\cite{SCH97} The fourth column lists the
(normalised) $GW$ correction to the density that results if the Hartree
potential is kept fixed at its LDA value. As one would expect, this unscreened
correction is larger. However, because contributions from all energies are
integrated over and we are looking at non-zero $\bG$ the screening effect is
much weaker than one might naively expect by scaling down with the dielectric
constant. The LDA structure factors of the pseudo valence charge are themselves
not very meaningful, as they depend to some extent on the specific
pseudopotential used. They are only listed to give an idea of the relative
magnitude of the $GW$ correction.

Table \ref{FFSI2} shows the structure factors of the total density of Si for
several reciprocal lattice vectors. The second column lists the LDA values and
the third the (normalised) $GW$ values. The biggest contribution comes
actually from the cores, as the comparison with the valence structure factors
in Table \ref{FFSI} shows, so that now the difference between the LDA and the
$GW$ structure factors looks even less significant. Because of the weight of
the core contribution the lattice constant has a crucial influence on the
structure factors. We have used for both the LDA and the $GW$ calculation the
experimental value of 5.43\AA. In the last column we list experimental static
(zero temperature) structure factors for comparison. They are the Fourier
transforms of a model fit\cite{DEU92,LZD93} to experimental data. The agreement
of both $GW$ and LDA with experiment is very good.

Table \ref{FFGE} gives the valence structure factors of germanium and the $GW$
correction to it. The raw $GW$ valence density integrates to 8.004 electrons
per unit cell (instead of 8), a charge violation of 0.05\%. The values shown
are for the normalised density and values for the unscreened correction are
given as in Si. Table \ref{FFGE2} compares the full LDA and $GW$ density
structure factors in bulk Ge with experimental static structure factors from
Ref.\ \onlinecite{LZD95} (fit C in their Table III). The agreement of both LDA
and $GW$ values with experiment is again very good. We have used a Ge lattice
constant of 5.66\AA\ in LDA and $GW$ calculations.

Note that the structure factors listed in the tables have to be multiplied by
the diamond lattice structure factors $S(\bG)$ to get the Fourier coefficients
of the charge density in the crystal as expressed by Eq.\ (\ref{RHOSOLID}).

Small though the differences between the LDA and $GW$ densities are, we can see
that for both Si and Ge the $GW$ charge in real space is slightly less
concentrated near the atom sites and moves a little into the interstitial and
bonding regions. As a trend this goes into the right direction since it is
known that the LDA tends to accumulate too much charge near the atoms. Figs.\
\ref{DIFFDENSSI} and \ref{DIFFDENSGE} show the charge density differences in Si
and Ge along the edge of a conventional cell between two atom sites.

We give a list of quasiparticle energies at several symmetry points for Si in
Table \ref{DELTAESI} and for Ge in Table \ref{DELTAEGE}. The second column
gives the values with the Hartree potential taken at LDA level and the third
with a Hartree potential which uses the $GW$ density. One can see that the
influence of $\Delta V_H$ on the quasi-particle energies is very small.

\section{Conclusion}

The results presented in the previous section show that in the materials we
have investigated the $GW$ approximation can for most practical purposes be
considered to be charge conserving. It could be demonstrated that the charge
density is very close to experimental values and also little different from the
density at LDA level. This gives support to the common practice of computing
the quasi-particle corrections to the LDA eigenvalues without adjusting the
Hartree potential. Adjusting the Hartree potential to the $GW$ density has only a
small effect on the quasi-particle energies.

\section{Appendix}

To show how we treat the high-energy tail of the integrand in Eq.\ (\ref{DRHO})
let us rewrite Eq.\ (\ref{DYSON}) in the form
\begin{equation}
G(i\omega) = F^{-1}(i\omega) G_0(i\omega),
\end{equation}
\begin{equation}
F(i\omega) = \bigl(1 - G_0(i\omega)\left[\Sigma_{xc}(i\omega) + \Delta V_H
                     - V_{xc} - \epsilon_0\right]\bigr).
\end{equation}

For large $\omega$, as $G_0$ decays as $1/\omega$, we can expand $F^{-1}$ in
powers of $G_0$, keeping in mind that $\Sigma$ itself contains one factor
$G_0$. To first order in $G_0$ then
\begin{equation}
\lim_{\omega \rightarrow -\infty } F^{-1}(i\omega)
= 1 + G_0(i\omega)\left[\Sigma_x + \Delta V_H - V_{xc} - \epsilon_0\right],
\end{equation}
where $\Sigma_x$ is the exchange part of the self-energy that does not itself
depend on the high energy tail of $G_0$,
\begin{equation}
\Sigma_x(i\omega) = G_0(0) V_c,
\end{equation}
where $V_c$ denotes the inter-electronic Coulomb interaction.

To this order of approximation then
\begin{eqnarray}
\lefteqn {
\lim_{\omega \rightarrow \infty} \Delta G(i\omega) } \nonumber \\
&& = G_0(i\omega)\left[\Sigma_x + \Delta V_H - V_{xc}\right]
G_0(i\omega).
\end{eqnarray}
After a few transformations we can find from this
\begin{eqnarray}
\lefteqn{
\lim_{\omega \rightarrow -\infty}
\mbox{Re} \Delta G(\br,\br;i\omega) = } \nonumber \\
&& \sum_\bk\sum_{nn'}\left\{
    \mbox{Re}\left[\Psi_{n\bk}(\br)\Psi_{n'\bk}(\br)
    P_{nn'}(\bk)\right] f^{(1)}_{nn'}(\omega;\bk) \right. \nonumber \\
&& \left. - \mbox{Re}\left[\Psi_{n\bk}(\br)\Psi_{n'\bk}(\br)
    P_{nn'}(\bk)\right] f^{(2)}_{nn'}(\omega;\bk)
\right\}
\end{eqnarray}
where
\begin{equation}
f^{(1)}_{nn'}(\omega;\bk) =
  \mbox{Re}\frac{1}{(\omega_{n\bk} - i\omega)(\omega_{n'\bk} - i\omega)},
\end{equation}
\begin{equation}
f^{(2)}_{nn'}(\omega;\bk) =
  \mbox{Im}\frac{1}{(\omega_{n\bk} - i\omega)(\omega_{n'\bk} - i\omega)},
\end{equation}
\begin{eqnarray}
\lefteqn{
P_{nn'}(\bk) = \sum_{\bG,\bG'}\Psi_{n\bk}(\bG)\left[\Sigma_x(\bk,\bG,\bG')
\right. }
\nonumber \\
&& \left. {}+ \Delta V_H(\bG - \bG') - V_{xc}(\bG - \bG')\right]
\Psi_{n'\bk}(\bG').
\end{eqnarray}

The advantage of this formulation is that the energy dependent parts can be
integrated analytically. We therefore split up the integration in Eq.\
(\ref{DRHO}) into two parts, the first of which is performed numerically up
to some suitably chosen value $\omega_0$ and the remainder of the integral from
$\omega_0$ to $-\infty$ is evaluated analytically by using the relationship
\begin{eqnarray}
\lefteqn{
\int_{-\infty}^{\omega_0} f^{(1)}_{nn'}(\bk)(\omega) d\omega }
\nonumber \\
&& = \left\{
\begin{array}{ll}
\frac{1}{\omega_{n\bk} - \omega_{n\bk'}}
\left[\arctan{\frac{\omega_0}{\omega_{n\bk'}}}
- \arctan{\frac{\omega_0}{\omega_{n\bk}}}\right]
& \mbox{if}\ \omega_{n\bk} \neq \omega_{n'\bk}
\nonumber \\
\frac{1}{\omega_{n\bk}}\frac{\omega_0/\omega_{n\bk}}{1 +
(\omega_0/\omega_{n\bk})^2}
& \mbox{if}\ \omega_{n\bk} = \omega_{n'\bk}
\nonumber
\end{array}
\right.
\end{eqnarray}
and
\begin{eqnarray}
\lefteqn{
\int_{-\infty}^{\omega_0} f^{(2)}_{nn'}(\bk)(\omega) d\omega }
\nonumber \\
&& = \left\{
\begin{array}{ll}
\frac{1}{2({\omega^>}^2 - {\omega^<}^2)}
\ln{\frac{\omega_0^2 + {\omega_{n\bk}^<}^2}{\omega_0^2 + {\omega_{n\bk}^>}^2}}
& \mbox{if}\ \omega_{n\bk} \neq \omega_{n'\bk}
\nonumber \\
- \frac{1}{2({\omega^>}^2 + {\omega^<}^2)}
& \mbox{if}\ \omega_{n\bk} = \omega_{n'\bk}
\nonumber
\end{array}
\right. ,
\end{eqnarray}
with $\omega^< = \min{\left(|\omega_{n\bk}|,|\omega_{n\bk'}|\right)}$,
$\omega^> = \max{\left(|\omega_{n\bk}|,|\omega_{n\bk'}|\right)}$


\begin{table}
\caption{\label{FFSI}
Structure factors (absolute values) of the pseudo valence charge
density of Si [e/atom] at LDA level and the correction from the $GW$
approximation with and without adjusting the Hartree potential.
}
\begin{tabular}{cddd}
$\bG$ & LDA valence & $\Delta\rho$ & $\Delta\rho$ \\
      &             & ($V_H$ $GW$)   & ($V_H$ LDA) \\
\hline
1 1 1 & 1.2487  & -0.0054 & -0.0076 \\
2 2 0 & 0.0323  & -0.0035 & -0.0043 \\
3 1 1 & 0.2429  & -0.0023 & -0.0023 \\
4 0 0 & 0.1877  &  0.0009 &  0.0012 \\
3 3 1 & 0.0601  &  0.0021 &  0.0023 \\
4 2 2 & 0.0685  &  0.0008 &  0.0010 \\
3 3 3 & 0.0755  & -0.0003 & -0.0001 \\
\end{tabular}
\end{table}

\begin{table}
\caption{\label{FFSI2}
Structure factors (absolute values) of the total charge density of Si [e/atom].
}
\begin{tabular}{cddd}
$\bG$ & LDA & $GW$ & \mbox{Exp.}\\
\hline
1 1 1 & 10.7210 & 10.7157 & 10.713 \\
2 2 0 &  8.6536 &  8.6501 &  8.655 \\
3 1 1 &  8.0205 &  8.0182 &  8.027 \\
4 0 0 &  7.4414 &  7.4423 &  7.454 \\
3 3 1 &  7.2256 &  7.2277 &  7.246 \\
4 2 2 &  6.6984 &  6.6992 &  6.712 \\
3 3 3 &  6.4086 &  6.4083 &  6.420 \\
\end{tabular}
\end{table}

\begin{table}
\caption{\label{FFGE}
Structure factors (absolute values) of the pseudo valence charge
density of Ge [e/atom] at LDA level and the correction from the $GW$
approximation with and without adjusting the Hartree potential.
}
\begin{tabular}{cddd}
$\bG$ & LDA valence & $\Delta\rho$ & $\Delta\rho$ \\
      &             & ($V_H$ $GW$)   & ($V_H$ LDA) \\
\hline
1 1 1 & 1.3181 & -0.0169 & -0.0239 \\
2 2 0 & 0.0035 & -0.0086 & -0.0107 \\
3 1 1 & 0.2201 & -0.0036 & -0.0033 \\
4 0 0 & 0.2052 &  0.0025 &  0.0037 \\
3 3 1 & 0.0974 &  0.0037 &  0.0040 \\
4 2 2 & 0.0865 &  0.0027 &  0.0033 \\
3 3 3 & 0.0726 &  0.0017 &  0.0023 \\
\end{tabular}
\end{table}

\begin{table}
\caption{\label{FFGE2}
Structure factors (absolute values) of the total charge density of Ge [e/atom].
}
\begin{tabular}{cddd}
$\bG$ & LDA & $GW$ & \mbox{Exp.} \\
\hline
1 1 1 & 27.5205 & 27.5036 & 27.453 \\
2 2 0 & 23.6819 & 23.6734 & 23.677 \\
3 1 1 & 22.1675 & 22.1639 & 22.138 \\
4 0 0 & 20.3205 & 20.3230 & 20.273 \\
3 3 1 & 19.4545 & 19.4582 & 19.509 \\
4 2 2 & 18.0361 & 18.0388 & 18.066 \\
3 3 3 & 17.2916 & 17.2939 & 17.315 \\
\end{tabular}
\end{table}

\begin{table}
\caption{\label{DELTAESI}
Quasi-particle energies of Si [eV].
}
\begin{tabular}{cdddd}
level & LDA & $GW$ + V$_H^{LDA}$ & $GW$ + V$_H^{GW}$ & \mbox{Exp.} \\
\hline
$\Gamma_{1c}$   &-11.88 &-11.91 &-11.91 & -12.50 \\
$\Gamma_{25'v}$ &  0.00 &  0.00 &  0.00 &   0.00 \\
$\Gamma_{15c}$  &  2.59 &  3.26 &  3.26 &   3.05 \\
$\Gamma_{2'c}$  &  3.26 &  4.05 &  4.03 &   4.1  \\[6pt]
$X_{1v}$        & -7.77 & -7.88 & -7.88 &  -8.18 \\
$X_{4v}$        & -2.81 & -2.92 & -2.92 &  -2.9  \\
$X_{1c}$        &  0.62 &  1.24 &  1.25 &   1.25 \\
$X_{4c}$        & 10.11 & 11.00 & 10.99 &  10.95 \\[6pt]
$L_{2'v}$       & -9.56 & -9.64 & -9.64 &  -9.3  \\
$L_{1v}$        & -6.95 & -7.09 & -7.08 &  -6.7  \\
$L_{3'v}$       & -1.16 & -1.22 & -1.22 &  -1.2  \\
$L_{1c}$        &  1.46 &  2.14 &  2.14 &   1.65 \\
$L_{3c}$        &  3.34 &  4.07 &  4.08 &   4.15 \\
$L_{2'c}$       &  7.73 &  8.34 &  8.36 &        \\[6pt]
Gap             &  0.49 &  1.10 &  1.11 &   1.17 \\
\end{tabular}
\end{table}

\begin{table}
\caption{\label{DELTAEGE}
Quasi-particle energies of Ge [eV].
}
\begin{tabular}{cdddd}
level & LDA & $GW$ + V$_H^{LDA}$ & $GW$ + V$_H^{GW}$ & \mbox{Exp.} \\
\hline
$\Gamma_{1c}$   &-12.63 &-12.79 &-12.79 & -12.60 \\
$\Gamma_{25'v}$ &  0.00 &  0.00 &  0.00 &   0.00 \\
$\Gamma_{2'c}$  &  0.01 &  0.66 &  0.63 &   0.89 \\
$\Gamma_{15c}$  &  2.61 &  3.11 &  3.14 &   3.21 \\[6pt]
$X_{1v}$        & -8.83 & -8.97 & -8.98 &  -9.3  \\
$X_{4v}$        & -2.98 & -3.11 & -3.09 &  -3.15 \\
$X_{1c}$        &  0.68 &  1.08 &  1.12 &   1.3  \\
$X_{3c}$        &  9.56 & 10.32 & 10.32 &        \\[6pt]
$L_{2'v}$       &-10.61 &-10.76 &-10.76 & -10.6  \\
$L_{1v}$        & -7.54 & -7.69 & -7.68 &  -7.7  \\
$L_{3'v}$       & -1.35 & -1.40 & -1.40 &  -1.4  \\
$L_{1c}$        &  0.14 &  0.64 &  0.66 &   0.74 \\
$L_{3c}$        &  3.77 &  4.29 &  4.32 &   4.3  \\
$L_{2'c}$       &  7.24 &  7.65 &  7.73 &   7.8  \\
\end{tabular}
\end{table}

\begin{figure}
\caption{\label{DIFFDENSSI}
Difference between the $GW$ and LDA charge densities\cite{CDENS} of Si (atomic
units) along the edge of a conventional cell between two atom sites. Distance
in Bohr radii. The $GW$ correction slightly increases the concentration of
electrons between the atoms.
}
\end{figure}

\begin{figure}
\caption{\label{DIFFDENSGE}
Difference between the $GW$ and LDA charge densities\cite{CDENS} of Ge (atomic
units) along the edge of a conventional cell between two atom sites. Distance
in Bohr radii. The $GW$ correction slightly increases the concentration of
electrons between the atoms.
}
\end{figure}

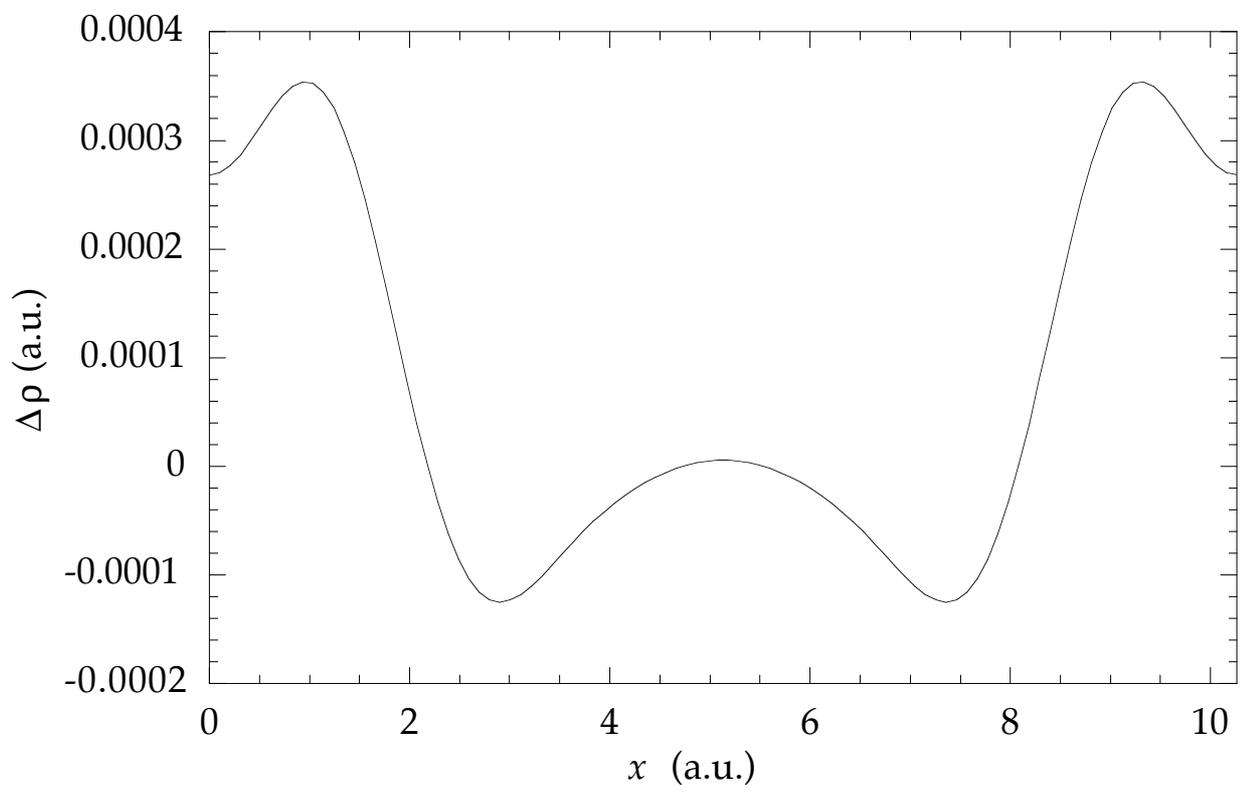
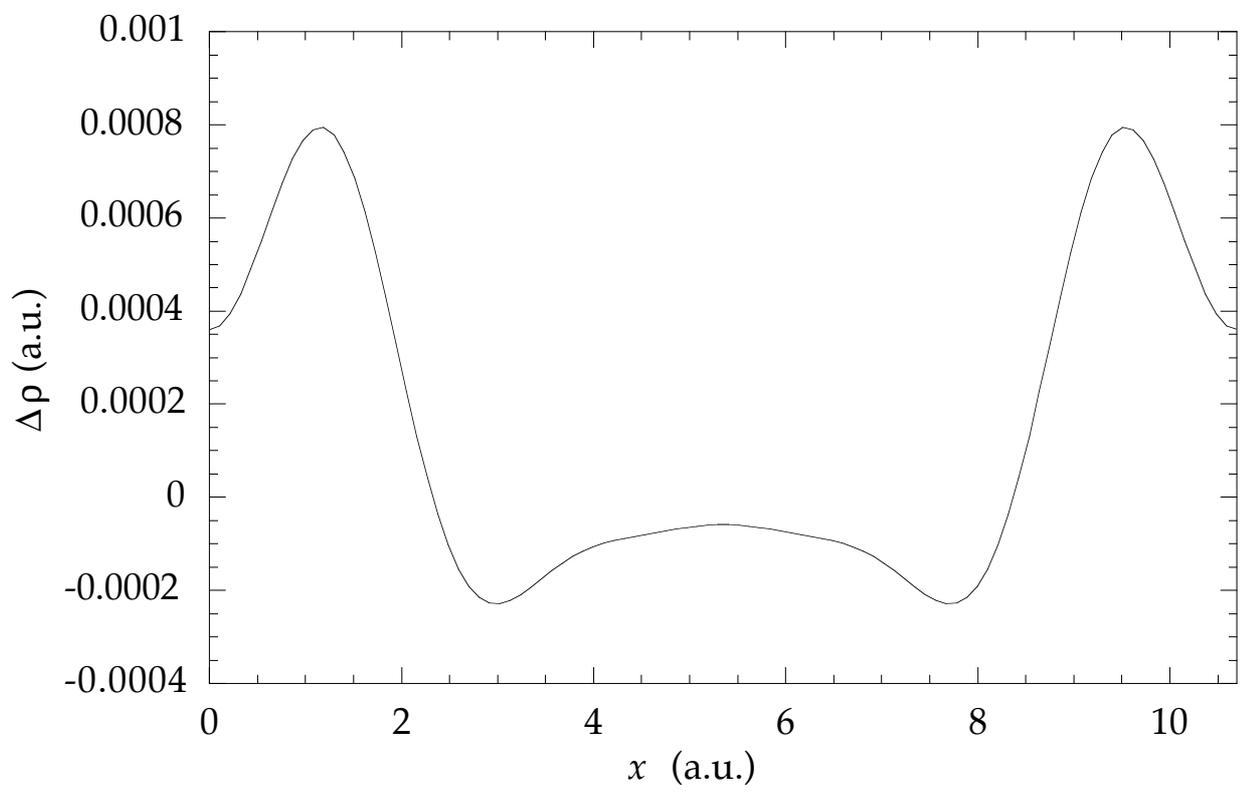
\end{document}